\documentclass[runningheads]{llncs}
\usepackage[T1]{fontenc}
\usepackage{graphicx,verbatim}
\usepackage{graphicx} 
\usepackage{amsmath}
\usepackage{amssymb}
\usepackage{amsfonts}
\usepackage{bm}
\usepackage{array}
\usepackage{mathtools}
\usepackage{url}
\usepackage[colorlinks=true, citecolor=green]{hyperref}
\usepackage{xcolor}
\usepackage{hyperref}
\usepackage{graphicx}
\usepackage{xcolor}
\usepackage{multirow}
\usepackage{graphicx}
\usepackage{float}  
\usepackage[table,dvipsnames]{xcolor}
\usepackage{dblfloatfix}
\usepackage{booktabs}
\begin{document}
	\title{\textbf{Endo-NeRF++}: Uncertainty-Aware Neural Rendering with Multi-Resolution Hash Encoding for Dynamic Surgical Scene Reconstruction}
	\titlerunning{\textbf{Endo-NeRF++}}
	
	\author{
		Gousia Habib\inst{1}\thanks{Corresponding author} \and
		Laura Ruotsalainen\inst{1}
	}
	\authorrunning{Habib et al.}
	\institute{
		\textsuperscript{1}Department of Computer Science, University of Helsinki, Finland
		\email{gousia.habib@helsinki.fi,laura.ruotsalainen@helsinki.fi}
	}

	\maketitle              
	\begin{center}
		\small
		\textbf{Project Page:} \\
		\url{https://github.com/gousiya26-I/EndoNerf-plusplus}
	\end{center}
	
	\begin{center}
		\textbf{Acknowledgements}
		\textit{This work was supported by the Academy of Finland Flagship Programme: Finnish Center for Artificial Intelligence (FCAI) and the Department of Computer Science, University of Helsinki. The authors acknowledge the research environment provided by ELLIS Institute Finland.}
	\end{center}
	
	\begin{abstract}
		Reconstructing dynamic surgical scenes is crucial for robot-assisted minimally invasive surgery; however, it continues to be difficult because of tissue deformation, occlusions, specular reflections, and restricted viewpoints. In this study, we introduce Endo-NeRF++, a neural rendering framework that accounts for uncertainty in the reconstruction of dynamic surgical scenes. Expanding on EndoNeRF, the suggested approach incorporates multi-resolution hash-grid encoding, temporal feature merging, and uncertainty-informed adaptive sampling to enhance reconstruction accuracy and temporal coherence in deformable endoscopic scenes.The multi-resolution hash-grid representation within the framework effectively captures both coarse and fine anatomical details, while temporal feature blending ensures stable reconstruction during tissue deformation and surgical tool occlusions. Additionally, uncertainty-driven adaptive sampling assigns more samples to uncertain areas to enhance rendering quality and geometric coherence. Experiments on robotic surgical video sequences demonstrate that the proposed uncertainty-guided adaptive sampling improves PSNR by up to 1.22\,dB (4.3\%), increases SSIM by up to 5.3\%, and reduces LPIPS by up to 55.1\% using EndoNerf dataset compared with the EndoNeRF baseline.
		\keywords{NeRFs \and Multiresolution Hash Grid \and Uncertainty Quantification \and Temporal Feature Blending \and 3D Reconstruction \and Robotic Surgery}
		
	\end{abstract}
	
	\section{Introduction}
	Reconstructing the surgical scene from stereoscopic video is particularly vital for robot-assisted minimally invasive surgery ~\cite{springer2026airobotics,cureus2024airobotics}, as it facilitates the retrieval of three-dimensional (3D) representations of the intraoperative environment. Three-dimensional (3D) reconstruction enhances spatial perception compared to conventional two-dimensional (2D) visualisation ~\cite{maierhein2017surgical}. This helps clinicians learn more about how tissues interact, how deep they are, and how they are arranged in the body ~\cite{mountney2010tissue}. Such a capability is necessary for making surgery more accurate and helping with more complex clinical workflows.\\
	The potential to produce precise and consistent 3D models of surgical environments has significant implications for a variety of downstream applications, including the automation of robotic surgery ~\cite{maierhein2017surgical}, augmented reality (AR)-enhanced procedures ~\cite{pratt2018hololens}, and virtual reality (VR)-driven training ~\cite{nicolau2011augmented}. Consequently, the reconstruction of a surgical scene is a critical component of the development of intelligent and beneficial surgical systems.\\
	Reconstructing dynamic surgical scenes is still very difficult, despite its significance. Non-rigid tissue deformation, specular reflections, occlusions, and low-texture areas are characteristics of endoscopic environments that make accurate depth estimation and 3D modeling challenging ~\cite{hirschmuller2008stereo}. Traditional pipelines typically employ stereo matching to estimate depth maps ~\cite{scharstein2002taxonomy,schoenberger2016sfm,newcombe2011kinectfusion} followed by RGB-D fusion techniques to reconstruct 3D geometry ~\cite{whelan2015elasticfusion,engel2018dso,mildenhall2020nerf,mueller2022instantngp}. However, these approaches often struggle to maintain robustness and accuracy under realistic surgical conditions.\\
	High-fidelity 3D reconstruction from multi-view images has been enabled by neural rendering methods, particularly Neural Radiance Fields (NeRF) ~\cite{mildenhall2020nerf,mueller2022instantngp}. Their dynamic extensions further model non-rigid scenes using deformation fields ~\cite{park2021nerfies,pumarola2021dnerf}. In the surgical domain, EndoNeRF ~\cite{wang2024endonerf}, EndoSurf ~\cite{zha2023endosurf} and NeRFscopy ~\cite{salortbenejam2026nerfscopy} incorporates deformation modeling, tool-aware masking, and geometry-aware rendering for stereo endoscopic video. However, these methods rely on single-network representations, limiting their ability to capture high-frequency details and maintain geometric consistency under rapid tissue deformation.\\
	Learning-based stereo depth estimation has achieved strong performance in recovering dense and accurate depth maps from image pairs ~\cite{lipson2021raftstereo,chang2018psmnet,zhang2019ganet}, providing robust geometric priors over traditional methods. However, their performance degrades in surgical settings due to specular highlights, low-texture regions, and dynamic tissue motion. Moreover, existing neural rendering approaches often underutilise stereo-derived geometry, leading to suboptimal reconstruction accuracy.\\
	Recent studies have examined uncertainty modelling in neural rendering and vision to improve robustness and enable adaptive computation ~\cite{yu2023uncertaintynerf,martinbrualla2021nerfwild,duan2026evidentialnerf}. Multi-resolution encodings, such as hash-grid representations, enhance high-frequency detail and facilitate optimisation ~\cite{mueller2022instantngp}. Nonetheless, the integration of stereo geometry, multi-resolution encoding, and uncertainty-aware sampling into a unified framework for deformable surgical scene reconstruction remains largely unexplored.\\
	To overcome these challenges, we propose a new method that combines stereo depth learning with multi-resolution hash encoding and uncertainty quantification with neural rendering to improve surgical scene reconstruction. Our method builds upon the EndoNeRF framework~\cite{wang2024endonerf}, extending its deformable neural rendering pipeline for surgical scene reconstruction. While EndoNeRF provides a strong foundation for modelling dynamic endoscopic scenes, it relies on a single MLP-based representation and fixed sampling strategy, which limits its ability to capture high-frequency details and adapt to uncertain regions.\\
	To address these limitations, our method introduces three key contributions.
	First, we replace the conventional MLP-based encoding with an ensemble of multi-resolution hash-grid encoders, enabling efficient learning of both coarse and fine geometric structures. Second, we incorporate a temporal feature blending module that improves consistency across frames under non-rigid motion and occlusions. Third, we propose an uncertainty-aware adaptive sampling strategy that dynamically allocates samples to regions with high predictive uncertainty, improving reconstruction accuracy and robustness. Together, these contributions enable more accurate, temporally consistent, and uncertainty-aware reconstruction of deformable surgical scenes while maintaining the overall EndoNeRF framework.
	\section{Related Work}
	The related work is organized into three categories: dynamic NeRF methods, endoscopic neural rendering approaches, and uncertainty-aware neural rendering techniques. \\
	\textbf{NeRFs and D-NeRFs:} The study concentrates on neural radiance fields and their dynamic expansions for the modelling of time-varying situations. Static NeRF methodologies~\cite{mildenhall2020nerf} provide high-quality renderings but are inherently confined to static environments, hence constraining their utility in practical applications. Dynamic NeRF methodologies ~\cite{pumarola2021dnerf,kirschstein2023nersemble,wang2024endonerf} seek to address this shortcoming by representing temporal fluctuations via deformation fields; yet, they entail considerable computing complexity and frequently encounter scalability challenges. Recent endeavours emphasise enhancing efficiency via alternative representations, including hash encodings and voxel-based techniques~\cite{mueller2022instantngp,luo2025hash,kim2024facthash}. Nevertheless, balancing computational efficiency, memory consumption, and reconstruction accuracy remains challenging, while the reliability of the reconstructed outputs is rarely quantified through explicit uncertainty estimation..\\
	\textbf{Endoscopic NeRFs:} Endoscopy-specific NeRF techniques further expose the constraints of general dynamic radiance field approaches in actual surgical settings. Early studies such as EndoNeRF~\cite{wang2022endonerf} have adapted dynamic NeRF formulations to deformable tissues. However, they remain computationally expensive and rely on implicit representations, which limit their geometric accuracy and downstream utility. EndoSurf~\cite{zha2023endosurf} and NeRFscopy~\cite{salort2026nerfscopy} are examples of later techniques that improve geometric modelling by employing explicit surface representations or structured deformations. However, they are heavily dependent on limited acquisition settings and depth priors. Although recent methods, such as Efficient EndoNeRF~\cite{wang2024endonerf} and Endo2DGS~\cite{endo2dgs2025}, aim to improve computational efficiency and reconstruction fidelity, reconstruction under complex non-rigid deformations, constrained viewpoints, and occlusions remains unreliable, while the associated uncertainty is rarely quantified.\\
	\textbf{Uncertainty-Aware NERF Sampling and Efficient Representations:} A third area of research is the development of efficient representations for neural radiance fields and uncertainty-aware modelling. Ensemble, Bayesian, and likelihood-based are the three primary categories of uncertainty quantification techniques that are currently in use. Although Bayesian approaches provide more detailed uncertainty estimates, they incur substantial computing burden as a result of sampling~\cite{duan2026evidential}. In contrast, likelihood-based models effectively capture aleatoric uncertainty but do not account for epistemic uncertainty. Ensemble-based approaches improve uncertainty estimation by incorporating model variance; however, they incur substantial memory and training expenses~\cite{sunderhauf2022dane}.\\
	Though they frequently rely on approximations or indirect cues like multi-view consistency, more recent methods seek to improve practicality by introducing post-hoc and training-free uncertainty estimation frameworks, such as Bayes' Rays~\cite{goli2024bayesrays} and WarpRF~\cite{safadoust2026warprf}. By shifting learning from MLPs to structured feature grids, effective representations like multi-resolution hash grids~\cite{mueller2022instantngp,luo2025hash} greatly speed up training and enhance expressivity.\\
	Nevertheless, the applicability of prior work in complex and deformable surgical environments is still restricted, as existing methods primarily address uncertainty estimation and efficient representations independently. The applicability would be improved by incorporating uncertainty into adaptive sampling strategies that are necessary for accurate reconstruction under dynamic and ambiguous conditions.
	\section{Overview of Proposed framework}
\begin{figure}[H]
		\centering
		\includegraphics[scale=0.34]{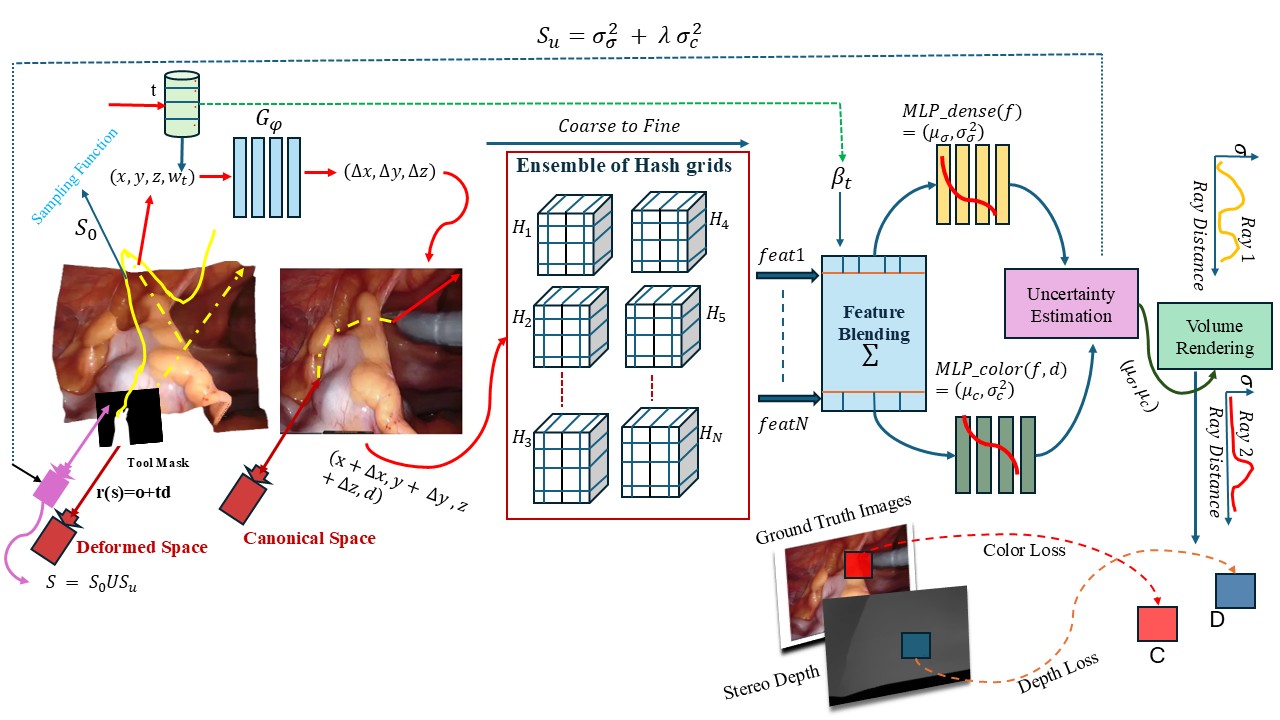}
		\caption{Proposed Framework}
		\label{fig:Fig1}
	\end{figure}

	Our approach expands upon the EndoNeRF architecture~\cite{wang2022endonerf}, which offers a robust framework for reconstructing deformable endoscopic scenes. Similar to EndoNeRF~\cite{wang2022endonerf}, we use: (i) a tool-aware canonicalization module to compensate for soft-tissue deformation; (ii) tool masks to avoid surgical instrument appearance contamination; (iii) depth cueing and geometry-aware rendering to stabilize the reconstruction; and (iv) a marching-based bidirectional mapping between canonical and deformed spaces.\\
	Despite its advantages, EndoNeRF's capacity to capture abrupt geometric distortions and delicate anatomical details is limited because it is based on a dynamic NeRF formulation that requires temporally smooth motion and relatively modest tissue deformation. Furthermore, EndoNeRF's ability to portray high-frequency fluctuations typical of surgical scenes is limited since it uses a single MLP to convey scene geometry and appearance. We present three important extensions to overcome these constraints.\\
	First, we replace the original MLP with \emph{an ensemble of multi-resolution hash-grid encoders}. These encoders provide better high-frequency spatial encoding, quick optimisation, and more accurate modelling of complex tissue structures. The hash grids' hierarchical structure makes it possible to capture both fine local information and coarse global shape, greatly enhancing reconstruction fidelity in areas experiencing fast deformation.\\
	Second, we incorporate a \emph{temporal feature blending module}, which fuses multi-scale features across time using learned temporal weights. This mechanism stabilizes the latent scene representation under tool-induced occlusions and non-rigid motion, and enables more temporally consistent appearance and geometry estimation.\\
	Third, EndoNeRF employs a fixed ray-sampling strategy that does not adjust sample allocation based on prediction confidence. In contrast, we propose an \emph{uncertainty-guided adaptive sampling} strategy that reallocates sampling density toward regions of high epistemic or aleatoric uncertainty. This allows the model to focus computation on geometrically ambiguous or rapidly deforming areas, leading to improved color and depth estimation.\\
	These three modifications (1) an ensemble of multi-resolution hash-grid encoders, (2) a temporal feature blending module, and (3) uncertainty-aware adaptive sampling form the core contributions of our approach. All other architectural components follow EndoNeRF’s original formulation.\\
	The proposed framework is illustrated in Figure  \ref{fig:Fig1}
	\section{Method}
	\subsection{Adaptive uncertainty-aware sampling and multi-resolution hash grids}
	Neural Radiance Fields (NeRFs) represent a scene as a continuous radiance field parameterized by a neural network:
	\begin{equation}
		F_{\theta} : (\mathbf{x},\mathbf{d}) \mapsto (\mathbf{c},\sigma),
		\label{eq:nerf_radiance_field}
	\end{equation}
	where $\mathbf{x} \in \mathbb{R}^{3}$ denotes a three-dimensional spatial location, $\mathbf{d}$ represents the viewing direction, $\mathbf{c}$ denotes the emitted color, and $\sigma$ is the volume density. Novel-view images are synthesized through volume rendering by integrating the predicted colors and densities along the corresponding camera rays. 
	We adopt an uncertainty-aware adaptive sampling strategy tailored for deformable surgical scenes. A ray is parameterised as
	\begin{equation}
		r(s) = \mathbf{o} + s \mathbf{d},
	\end{equation}
	where $\mathbf{o}$ denotes the camera origin, $\mathbf{d}$ is the unit direction of the ray, and $s$ represents the distance along the ray,
	from which initial samples $\mathbf{x}_j = \mathbf{r}(s_j)$ are drawn. 
	During an initial warm-up phase, we follow the sampling strategy of EndoNeRF~\cite{wang2022endonerf}, allowing the deformation network and appearance fields to stabilise before uncertainty estimates become reliable.\\
	After the warm-up, which is an initial training stage where the model first learns the basic scene geometry before estimating deformations, reducing optimisation ambiguity and improving convergence, we switch to our uncertainty-guided adaptive sampling, which redistributes samples toward regions exhibiting high predictive uncertainty.\\
	Each sampled point lies in the deformed scene space $S=S_0\cup S_u$ and is mapped to a canonical configuration using the deformation network $G_{\phi}$:
	\begin{equation}
		\Delta \mathbf{x}_j = G_\varphi(\mathbf{x}_j, w_t), \qquad 
		\mathbf{x}_j^0 = \mathbf{x}_j + \Delta\mathbf{x}_j,
	\end{equation}
	where $w_t$ encodes surgical tool motion. The canonical points $\mathbf{x}_j^0$ query a coarse-to-fine ensemble of multi-resolution hash grids $\{H_i\}$, producing features $\mathbf{f}_j^{(i)} = H_i(\mathbf{x}_j^0)$, which are blended using temporal weights:
	\begin{equation}
		\mathbf{f}_j = \sum_{i} \beta_t^{(i)} \mathbf{f}_j^{(i)},
	\end{equation}
	where the temporal blending weights $\{\beta_t^{(i)}\}$ are normalized such that $\sum_i \beta_t^{(i)} = 1$.
	Two MLP heads regress predictive means and log-variances for density and color:
	\begin{align}
		(\mu_{\sigma,j}, \ell_{\sigma,j}) &= \mathrm{MLP}_{\text{dense}}(\mathbf{f}_j), \\
		(\boldsymbol{\mu}_{c,j}, \ell_{c,j}) &= \mathrm{MLP}_{\text{color}}(\mathbf{f}_j, \mathbf{d}).
	\end{align}
	To capture epistemic uncertainty, we enable dropout during both training and inference and perform $K$ stochastic forward passes. For each sampled point $\mathbf{x}_j$, this yields:
	\begin{equation}
		\{ \mu_{\sigma,j}^{(k)}, \sigma_{\sigma,j}^{2(k)}, \boldsymbol{\mu}_{c,j}^{(k)}, \sigma_{c,j}^{2(k)} \}_{k=1}^{K},
	\end{equation}
	where $\sigma^{2(k)} = \mathrm{softplus}(\ell^{(k)}) + \epsilon$, with a small constant $\epsilon$ added for numerical stability.
	The predictive means are computed as:
	\begin{equation}
		\bar{\mu}_{\sigma,j} = \frac{1}{K} \sum_{k=1}^{K} \mu_{\sigma,j}^{(k)}, \quad
		\bar{\boldsymbol{\mu}}_{c,j} = \frac{1}{K} \sum_{k=1}^{K} \boldsymbol{\mu}_{c,j}^{(k)}.
	\end{equation}
	
	Aleatoric uncertainty is given by:
	\begin{equation}
		\sigma_{\text{alea},\sigma,j}^{2} = \frac{1}{K} \sum_{k=1}^{K} \sigma_{\sigma,j}^{2(k)}, \quad
		\sigma_{\text{alea},c,j}^{2} = \frac{1}{K} \sum_{k=1}^{K} \sigma_{c,j}^{2(k)}.
	\end{equation}
	
	Epistemic uncertainty is estimated as:
	\begin{equation}
		\sigma_{\text{epi},\sigma,j}^{2} = \frac{1}{K} \sum_{k=1}^{K}
		\left(\mu_{\sigma,j}^{(k)} - \bar{\mu}_{\sigma,j}\right)^2,
	\end{equation}
	\begin{equation}
		\sigma_{\text{epi},c,j}^{2} = \frac{1}{K} \sum_{k=1}^{K}
		\left(\boldsymbol{\mu}_{c,j}^{(k)} - \bar{\boldsymbol{\mu}}_{c,j}\right)^2.
	\end{equation}
	The total predictive uncertainty is:
	\begin{equation}
		\sigma_{\sigma,j}^{2} =
		\sigma_{\text{alea},\sigma,j}^{2} + \sigma_{\text{epi},\sigma,j}^{2}, \quad
		\sigma_{c,j}^{2} =
		\sigma_{\text{alea},c,j}^{2} + \sigma_{\text{epi},c,j}^{2}.
	\end{equation}
	Before combining these uncertainty components, each variance is min--max normalized along the corresponding camera ray  to ensure comparable scaling. The normalized variance is denoted by $\tilde{\sigma}^{2}$:
	\begin{equation}
		\tilde{\sigma}_{q,j}^{2}
		=
		\frac{\sigma_{q,j}^{2}-\min_{k}\sigma_{q,k}^{2}}
		{\max_{k}\sigma_{q,k}^{2}-\min_{k}\sigma_{q,k}^{2}+\epsilon},
		\qquad q\in\{\sigma,c\},
		\label{eq:uncertainty_normalization}
	\end{equation}
	where $k$ indexes samples along the ray and $\epsilon$ is a small positive constant introduced for numerical stability.
	A per-sample uncertainty score combines geometric (density) and photometric (color) uncertainty:
	\begin{equation} 
		S^{(\mathrm{sample})}_{u,j}
		= \tilde{\sigma}_{\sigma,j}^{2} + \lambda\,\tilde{\sigma}_{c,j}^{2},
		\label{eq:sample_uncertainty}
	\end{equation}
	Samples with higher uncertainty are allocated additional points proportionally to their normalized uncertainty score, while a minimum allocation is enforced to prevent sample collapse.
	During volume rendering, the predicted density is converted into an opacity value for each sampled point along a camera ray. The opacity, $\alpha_j$, determines how much light is absorbed or emitted at that location, whereas the accumulated transmittance, $T_j$, represents the fraction of light that reaches the point without being occluded by preceding samples. The final pixel colour is computed by accumulating the colors of all sampled points, weighted by their corresponding opacity and transmittance values.
	\begin{equation}
		\alpha_j = 1 - \exp(-\bar{\mu}_{\sigma,j}\Delta s_j), \qquad
		T_j = \exp\!\left(-\sum_{k<j} \bar{\mu}_{\sigma,k}\Delta s_k\right).
	\end{equation}
	Predicted color and depth are obtained via:
	\begin{equation}
		\hat{\mathbf{C}} = \sum_{j} T_j \alpha_j\,\bar{\boldsymbol{\mu}}_{c,j}, 
		\qquad
		\hat{D} = \sum_{j} T_j \alpha_j\, s_j.
	\end{equation}
	Although the rendered color and depth depend only on the predictive means $(\bar{\mu}_\sigma, \bar{\mu}_c)$, uncertainty is explicitly modeled through both aleatoric and epistemic components. Aleatoric uncertainty arises from the predicted variances, while epistemic uncertainty is estimated via Monte Carlo dropout as the variance across multiple stochastic forward passes. Density uncertainty is not directly used in rendering but contributes to uncertainty estimation and adaptive sampling.\\
	During training, a single stochastic forward pass is used for efficiency, whereas multiple passes are employed during inference to estimate epistemic uncertainty. The combined predictive uncertainty guides adaptive sampling, allocating more samples to regions with high data or model uncertainty. This adaptive sampling alters the distribution of samples along each ray, thereby influencing the estimated transmittance and opacity fields $(T_j, \alpha_j)$. As a result, depth estimation is indirectly influenced by uncertainty, even though rendering itself relies only on the predictive mean density.
	A ray-level aggregated uncertainty is computed as
	\begin{equation} \label{Eq-ray}
		S^{(\mathrm{ray})}_u =
		\frac{\sum_j T_j \alpha_j \left( \sigma^2_{\sigma,j} + \lambda \sigma^2_{c,j} \right)}
		{\sum_j T_j \alpha_j}.
	\end{equation} 
	The \textbf{Equation \ref{Eq-ray}} gives to an opacity-weighted average of per-sample uncertainty along the ray. This aggregated score drives final sample redistribution after the warm-up phase.
	\paragraph{Training Objective.}
	We supervise prediction using a negative log-likelihood (NLL) loss:
	\begin{equation}
		\mathcal{L}^{\mathrm{NLL}}_{\mathrm{rgb}} =
		\sum_{\text{rays}} \sum_j w_j
		\left(
		\frac{\| \bar{\mu}_{c,j} - \mathbf{C}_{\mathrm{gt}} \|_2^2}{2\sigma^2_{c,j}} 
		+ \frac{1}{2}\log \sigma^2_{c,j}
		\right),
	\end{equation}
	where $w_j = T_j \alpha_j$, $\bar{\mu}_{c,j}$ and $\sigma^2_{c,j}$ denote the predicted mean color and variance for the $j$-th sample, and $\mathbf{C}_{\mathrm{gt}}$ is the ground truth RGB color corresponding to the ray.\\
	Depth is supervised via an $\ell_1$ loss or an NLL variant when depth variance is predicted. Additionally, we apply a small variance regularizer to prevent variance inflation:
	\begin{equation}
		\mathcal{R}(\sigma) = \sum_{j} \log (1 + \sigma_{c,j}^{2}).
	\end{equation}
	The full training objective is:
	\begin{equation}
		\mathcal{L} = \mathcal{L}^{\mathrm{NLL}}_{\mathrm{rgb}} 
		+ \lambda_d \mathcal{L}_{\mathrm{depth}} 
		+ \eta \mathcal{R}(\sigma).
	\end{equation}
	where $\mathcal{L}$ is the total loss, $\mathcal{L}_{\mathrm{NLL}}^{\mathrm{rgb}}$ is the RGB negative log-likelihood loss, $\mathcal{L}_{\mathrm{depth}}$ is the depth loss, and $\mathcal{R}(\sigma)$ is the regularization term applied to the predicted density $\sigma$. The coefficients $\lambda_d$ and $\eta$ control the contributions of the depth loss and density regularization, respectively.\\
	As illustrated in Figure~\ref{fig:Fig1}, the predictive means $(\bar{\mu}_{\sigma}, \bar{\mu}_{c})$ are used for rendering, while the predicted variances $(\sigma_{\sigma}^{2}, \sigma_{c}^{2})$ influence uncertainty estimation and adaptive sampling. Consequently, depth is indirectly influenced by uncertainty through adaptive sampling, even though rendering relies only on the predictive mean density.X
	\section{Experiments}
	\subsection{Settings}
	\textbf{Dataset and Evaluation Metrics}:
	We assess our approach using stereo endoscopic video clips from the proprietary DaVinci robotic prostatectomy dataset presented in EndoNeRF~\cite{wang2022endonerf}. Owing to restricted data availability, we employ a sparse subset of 2 video sequences that encompass tissue pulling and cutting scenarios, totalling roughly 250 frames. Each video clip lasts 4--8\,s with 15fps seconds and is recorded from stereo cameras at a single viewpoint, demonstrating difficult conditions including non-rigid deformation and tool occlusion.\\
	Although the dataset size is diminished, these sequences exemplify standard soft-tissue interactions observed in robotic surgery, encompassing deformation and surgical manipulation. We utilise the cutting-edge surgical scene reconstruction technique EndoNeRF~\cite{wang2022endonerf} and DSSR~\cite{long2021edssr} as a robust baseline for comparison.\\
	To further evaluate the generalizability of our approach, we additionally conduct experiments on the \textbf{StereoMIS} dataset~\cite{hayoz2023pose}, which provides stereo endoscopic videos and camera kinematics acquired using the da Vinci Xi surgical robot. The dataset contains in-vivo porcine surgical sequences exhibiting realistic challenges such as tissue deformation and instrument motion. We use the\textbf{\textit{P2\_2}} and \textbf{\textit{P2\_6}}sequences, selected due to the availability of segmentation masks and their relatively short durations, enabling efficient evaluation on a dataset distinct from EndoNeRF.\\
	We visually assess reconstructed point clouds for qualitative evaluation, comparing geometric and textural fidelity among various approaches.\\
	We perform ablation studies to assess the contribution of each element in our system, including the multi-resolution hash-grid encoding, temporal feature blending module, and uncertainty-aware adaptive sampling. We conducted ablation study of our model by substituting the hash-grid encoder with a conventional MLP representation, eliminating the temporal feature blending mechanism, and replacing the uncertainty-guided sampling strategy with a uniform sampling method akin to EndoNeRF~\cite{wang2022endonerf}.\\
	\textbf{Metrics:} Following the evaluation presented in ~\cite{long2021edssr} and prior work in neural rendering, we adopt standard photometric metrics Peak Signal-to-Noise Ratio \textbf{PSNR}, Structural Similarity Index Measure \textbf{SSIM}, and Learned Perceptual Image Patch Similarity \textbf{LPIPS} for quantitative evaluation.\\
	\textbf{Baselines: }To evaluate the efficacy of our methodology, we compare it with two cutting-edge techniques: \textbf{EndoNeRF}~\cite{wang2022endonerf} and 
	\textbf{E-DSSR}~\cite{long2021edssr}. EndoNeRF utilises neural radiance fields to describe dynamic endoscopic environments, whilst E-DSSR implements transformer-based stereoscopic depth estimation for effective surgical scene reconstruction.
	\subsection{Implementation Details}
	All experiments were performed on the Supercomputer using the AI GPU partition. Each GPU node is equipped with two Intel Xeon Gold 6230 (Cascade Lake) processors (40 CPU cores in total), four NVIDIA Tesla V100 GPUs with 32\,GB HBM2 memory each interconnected via NVLink, 384\,GB system memory, and 3.6\,TB of local NVMe storage.\\
	Here, $\xi$ denotes the width of the Gaussian transfer function used to concentrate sampled points around the stereo-estimated depth. Depth loss weight, depth refinement interval, and exponential moving average coefficient are set to  $\lambda_{\text{depth}}=1$, $K=4000$, and $\alpha=0.1$, respectively. Endoscope intrinsics are obtained through camera calibration, tool masks are manually annotated, and coarse stereo depth maps are generated using the pretrained STTR-light model. Multi-resolution hash encoding employs hash resolutions of $\{8,16,32\}$, scaling factors of $\{1.4,1.7,1.9\}$, 16 hash levels and 8 feature channels per level for both the pulling tissue and cutting tissue sequences. Each model is trained for 50K iterations using an uncertainty loss weight of $\lambda_{\text{unc}}=0.02$ for cutting tissues and 0.03 for pulling tissues. After an initial warm-up phase (5k iterations), adaptive ray sampling selects 30\% uncertainty-guided rays and 70\% uniformly sampled rays. All remaining training and rendering hyperparameters follow the original EndoNeRF settings.
	\subsection{Multi-Resolution Hash-Grid Encoding and Temporal Feature Blending}
	We initially compare our approach with the original MLP-based representation utilised in EndoNeRF in order to assess the efficacy of the suggested multi-resolution hash-grid encoding and temporal feature mixing module. The basic reconstruction metrics PSNR, SSIM, and LPIPS are used to evaluate both the tissue pulling and tissue cutting sequences. The ability of the suggested representation to capture fine anatomical structures, maintain geometric consistency, and preserve temporal coherence under non-rigid tissue deformation and surgical tool occlusions is demonstrated through qualitative comparisons in addition to the quantitative evaluation.
	\begin{table}[H]
		\caption{Quantitative comparison of reconstruction results on pulling and cutting sequences.}
		\label{tab:hashgrid_results}
		\resizebox{\textwidth}{!}{
			\begin{tabular}{|l|llc|llc|}
				\hline
				\multicolumn{1}{|c|}{\multirow{2}{*}{Method}} & \multicolumn{3}{c|}{\textbf{Pulling Tissues}}                          & \multicolumn{3}{c|}{\textbf{Cutting Tissues}}                          \\ \cline{2-7} 
				\multicolumn{1}{|c|}{}                        & \multicolumn{1}{c|}{\textbf{PSNR}$\textcolor{green!70!black}{\uparrow}$} & \multicolumn{1}{c|}{\textbf{SSIM}$\textcolor{green!70!black}{\uparrow}$} & \textbf{LPIPS}$\textcolor{red}{\downarrow}$ & \multicolumn{1}{c|}{\textbf{PSNR}$\textcolor{green!70!black}{\uparrow}$} & \multicolumn{1}{c|}{\textbf{SSIM}$\textcolor{green!70!black}{\uparrow}$} & \textbf{LPIPS}$\textcolor{red}{\downarrow}$ \\ \hline
				EndoNeRF~\cite{wang2022endonerf}                                      & \multicolumn{1}{l|}{$28.32 \pm 0.503$}     & \multicolumn{1}{l|}{$0.912 \pm 0.015$}     &   $0.107 \pm 0.019$    & \multicolumn{1}{l|}{$26.58 \pm 0.52$}     & \multicolumn{1}{l|}{$0.884\pm 0.012$}     & $0.13\pm 0.02$      \\ \hline
				EndoNeRF++ \textbf{(ours)}                                   & \multicolumn{1}{l|}{\textbf{$29.08\pm 0.199$}}     & \multicolumn{1}{l|}{\textbf{$0.949\pm 0.006$}}     & \textbf{$0.052\pm 0.008$}      & \multicolumn{1}{l|}{\textbf{$27.334\pm 0.391$}}     & \multicolumn{1}{l|}{\textbf{$0.93\pm 0.009$}}     &   \textbf{$0.076\pm 0.008$}    \\ \hline
		\end{tabular} }
	\end{table}
	\subsubsection{Quantitative Results} The quantitative comparison between the original EndoNeRF framework and the suggested multi-resolution hash-grid encoding with temporal feature blending on the tissue pulling and tissue cutting sequences is shown in Table~\ref{tab:hashgrid_results}. PSNR, SSIM, and LPIPS are used to assess performance; better reconstruction quality is indicated by higher PSNR and SSIM and lower LPIPS. The outcomes show how well the suggested representation works to increase reconstruction fidelity under difficult surgical scene deformations. \\
	The suggested EndoNeRF++ consistently beats EndoNeRF on both tissue pulling and tissue cutting sequences, as demonstrated in Table~\ref{tab:hashgrid_results}.\\
	Our approach increases SSIM from 0.912 to 0.949, decreases LPIPS from 0.107 to 0.052, and increases PSNR from 28.32 dB to 29.08 dB on the pulling sequence. The suggested approach also produces higher PSNR (27.33 dB), higher SSIM (0.930), and lower LPIPS (0.076) on the cutting sequence, indicating that it may rebuild highly deformable surgical scenes with better geometric and photometric integrity.
	\begin{figure}[H]
		\centering
		\includegraphics[width=\textwidth]{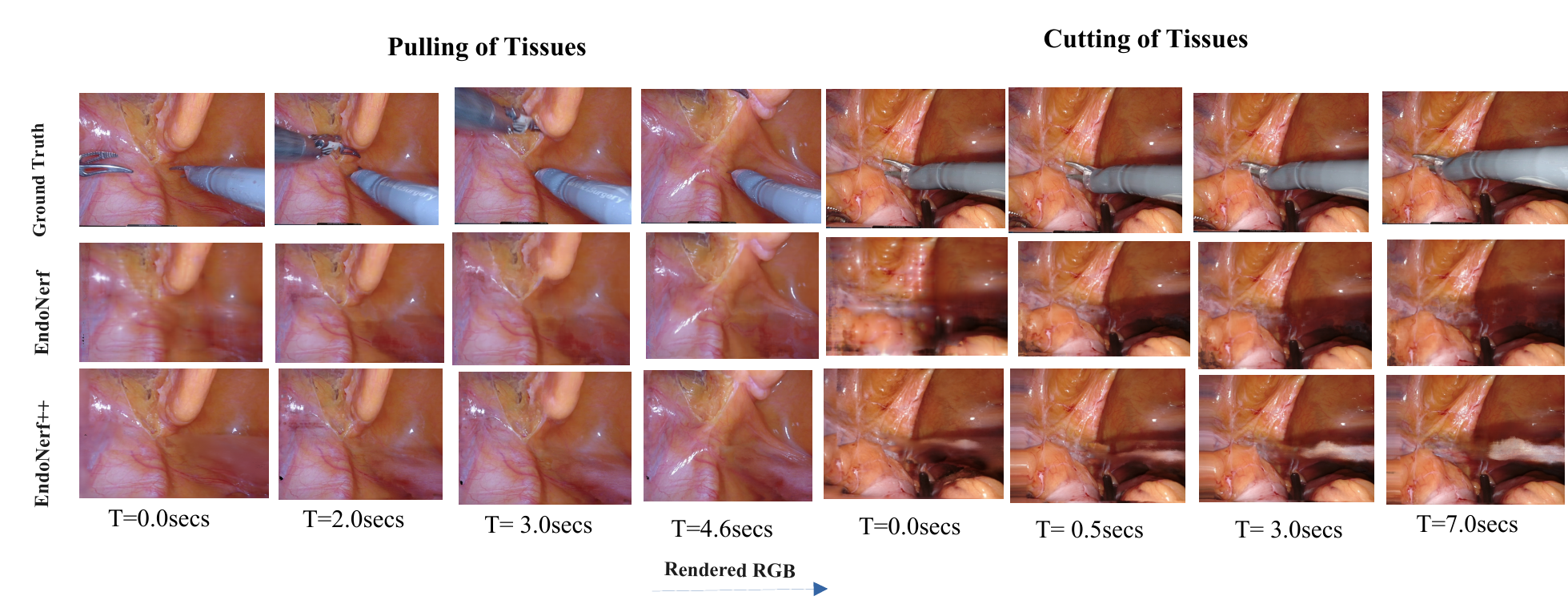}
		\caption{Qualitative comparison of reconstruction results on representative
			frames from the tissue pulling and tissue cutting sequences.}
		\label{fig:qualitative_results}
	\end{figure}
	\subsubsection{Qualitative Results} For both the tissue pulling and tissue cutting sequences, qualitative comparisons are shown in Fig.~\ref{fig:qualitative_results} to support the quantitative assessment. We compare the associated ground-truth images with the reconstructed images produced by EndoNeRF, E-DSSR, and the suggested EndoNeRF++. The visual outcomes in Fig.~\ref{fig:qualitative_results} show how well the suggested temporal feature mixing and multi-resolution hash-grid encoding work to rebuild dynamic surgical scenes with better geometric consistency, more precise anatomical features, and fewer rendering artefacts.
	\subsection{Multi-Resolution Hash Encoding with Uncertainty Estimation}
	After demonstrating the improved reconstruction performance of the proposed multi-resolution hash-grid encoding, we evaluate its ability to estimate predictive uncertainty in dynamic surgical scenes. Reliable uncertainty estimation is essential for identifying regions with low prediction confidence caused by occlusions, complex tissue deformations, and rapid camera motion. To achieve this, we augment the proposed EndoNeRF++ framework with an uncertainty prediction branch that jointly learns scene appearance and per-pixel predictive uncertainty during neural rendering. 
	\begin{table}[H]
		\centering
		\caption{Quantitative comparison of EndoNeRF++. Reconstruction results are reported as mean $\pm$ standard deviation $\Delta$, and uncertainty results as mean, standard deviation, best correlation, and AUSE.}
		\label{tab:uncertainty_results}
		\small
		\setlength{\tabcolsep}{4pt}
		\textbf{Reconstruction Performance}
		\vspace{2mm}
		
		\begin{tabular}{|l|c|c|c|}
			\hline
			\textbf{Sequence} &
			\textbf{PSNR} \textcolor{ForestGreen}{$\uparrow$} &
			\textbf{SSIM} \textcolor{ForestGreen}{$\uparrow$} &
			\textbf{LPIPS} \textcolor{red}{$\downarrow$} \\
			\hline
			Pulling &
			29.256$\pm$0.326 ($\uparrow$0.176) &
			0.943$\pm$0.008 ($\downarrow$0.006) &
			0.055$\pm$0.009 ($\uparrow$0.003) \\
			\hline
			Cutting &
			27.356$\pm$0.141 ($\uparrow$0.022) &
			0.931$\pm$0.012 ($\uparrow$0.001) &
			0.074$\pm$0.005 ($\downarrow$0.002) \\
			\hline
		\end{tabular}
		
		\vspace{3mm}
		\textbf{Uncertainty Estimation}
		\vspace{2mm}
		
		\begin{tabular}{|l|c|c|c|c|c|}
			\hline
			\textbf{Sequence} &
			\multicolumn{2}{|c|}{\textbf{Pearson} \textcolor{ForestGreen}{$\uparrow$}} &
			\multicolumn{2}{|c|}{\textbf{Spearman} \textcolor{ForestGreen}{$\uparrow$}} &
			\textbf{AUSE} \textcolor{red}{$\downarrow$} \\
			\cline{2-5}
			& \textbf{Mean$\pm$Std} & \textbf{Best} &
			\textbf{Mean$\pm$Std} & \textbf{Best} & \\
			\hline
			Pulling &
			0.201$\pm$0.0.051 & 0.302 & 
			0.280$\pm$0.075 & 0.465 &
			0.698 \\
			\hline 
			Cutting &
			0.380$\pm$0.034 & \textbf{0.511} &
			0.284$\pm$0.047 & \textbf{0.475} &
			0.487 \\
			\hline
		\end{tabular}
	\end{table}
	\subsubsection{Quantitative Results} The quantitative uncertainty estimation results are summarized in Table~\ref{tab:uncertainty_results}. The proposed method achieves a mean Pearson correlation of 0.380, a mean Spearman correlation of 0.284, and an AUSE of 0.487 for cutting tissues sequence and Pearson correlation of 0.201 and Spearman correlation of 0.280 and with an AUSE as 0.698. These results demonstrate a consistent positive association between the predicted uncertainty and reconstruction error, indicating that the proposed uncertainty estimates provide informative signals for identifying less reliable reconstructions while maintaining competitive reconstruction performance.\\
	The Cutting sequence exhibits a higher Pearson than Spearman correlation, indicating a stronger linear relationship between predicted uncertainty and reconstruction error. In contrast, the Pulling sequence shows a slightly higher Spearman correlation, suggesting better preservation of the relative error ranking. This difference is likely due to the varying motion patterns and tissue deformations across the two sequences.
	\begin{figure}[H]
		\centering
		\includegraphics[width=\textwidth]{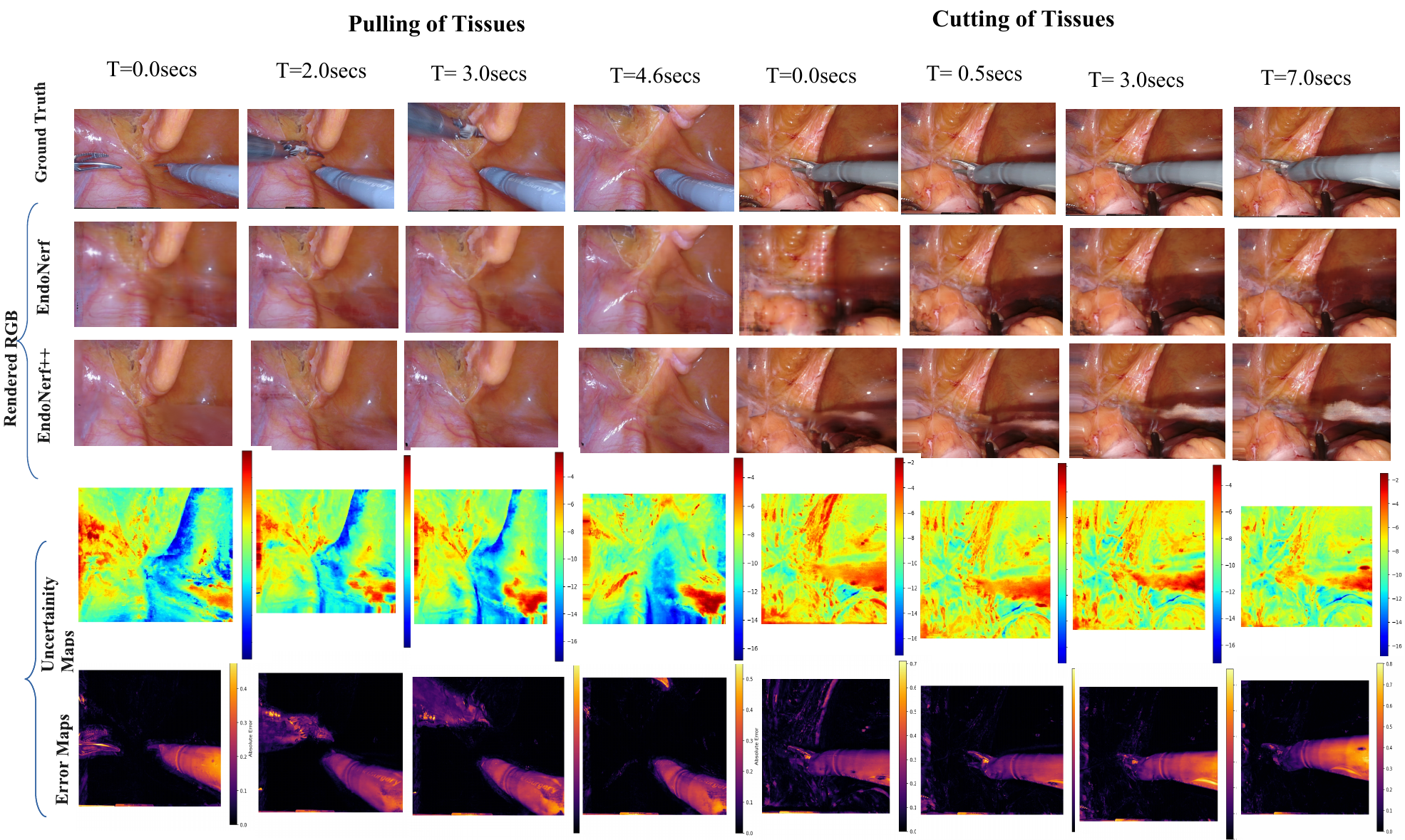}
		\caption{Qualitative comparison of EndoNeRF and EndoNeRF++ on representative frames. Uncertainty maps reflect reconstruction errors}
		\label{fig:uncertainty_results}
	\end{figure}
	These findings show that places with higher reconstruction error are consistently highlighted by the anticipated uncertainty, yielding significant confidence estimates without sacrificing reconstruction quality. The suggested approach shows that uncertainty estimation can be included without appreciably impairing reconstruction performance by achieving better PSNR while keeping comparable SSIM and LPIPS when compared to the baseline EndoNeRF++. 
	\subsubsection{Qualitative Results:} The reconstructed images, uncertainty maps, and reconstruction error maps for representative frames from the Pulling and Cutting Tissues sequences are qualitatively compared in Figure~\ref{fig:uncertainty_results}. Compared to the baseline EndoNeRF, the suggested EndoNeRF++ produces reconstructions that are visually closer to the ground truth. Furthermore, the efficiency of the proposed uncertainty estimating module is demonstrated by the tight correspondence between regions of high anticipated uncertainty and regions of high reconstruction error, especially near surgical tools, tissue borders, and areas enduring significant deformations.\\
	Taken together,the quantitative and qualitative findings show that the suggested uncertainty estimation module offers trustworthy confidence estimates that have a strong correlation with reconstruction mistakes. The following section explores the application of these uncertainty predictions for uncertainty-guided adaptive sampling.
	\subsection{Uncertainty-Guided Adaptive Sampling}
	An important question remains:\textbf{\textit{ Can the proposed uncertainty estimates be effectively exploited to improve neural rendering? }}To answer this, we evaluate the proposed uncertainty-guided adaptive sampling strategy, which prioritizes rays with higher predicted uncertainty during training. The quantitative and qualitative results are summarized in Table~\ref{tab:adaptive_results} and Figure ~\ref{fig:adaptive_results}, respectively.
	\begin{table}[H]
		\centering
		\caption{Performance of uncertainty-guided adaptive sampling compared with the baseline. Values are reported as mean$\pm$standard deviation across all test frames.}
		\label{tab:adaptive_results}
		\small
		\setlength{\tabcolsep}{4pt}
		\textbf{Reconstruction Performance}
		\vspace{2mm}
		
		\begin{tabular}{|l|c|c|c|}
			\hline
			\textbf{Sequence} &
			\textbf{PSNR} \textcolor{ForestGreen}{$\uparrow$} &
			\textbf{SSIM} \textcolor{ForestGreen}{$\uparrow$} &
			\textbf{LPIPS} \textcolor{red}{$\downarrow$} \\
			\hline
			Pulling &
			29.256$\pm$0.326 ($\uparrow$0.176) &
			0.943$\pm$0.008 ($\downarrow$0.006) &
			0.055$\pm$0.009 (+0.003) \\
			\hline
			Cutting &
			27.356$\pm$0.141 ($\uparrow$0.016) &
			0.931$\pm$0.012 ($\downarrow$0.008) &
			0.074$\pm$0.005 (+0.011) \\
			\hline
		\end{tabular}
		
		\vspace{3mm}
		
		\textbf{Uncertainty Estimation}
		\vspace{2mm}
		
		\begin{tabular}{|l|c|c|c|c|c|}
			\hline
			\textbf{Sequence} &
			\multicolumn{2}{|c|}{\textbf{Pearson} \textcolor{ForestGreen}{$\uparrow$}} &
			\multicolumn{2}{|c|}{\textbf{Spearman} \textcolor{ForestGreen}{$\uparrow$}} &
			\textbf{AUSE} \textcolor{red}{$\downarrow$} \\
			\cline{2-5}
			& \textbf{Mean$\pm$Std} & \textbf{Best} &
			\textbf{Mean$\pm$Std} & \textbf{Best} &
			\\
			\hline
			Pulling &
			0.201$\pm$0.0390 & \textbf{0.302} &
			0.6050$\pm$0.0670 & \textbf{0.669} &
			0.317\\
			\hline
			Cutting &
			0.242$\pm$0.038 & \textbf{0.321} &
			0.556$\pm$0.022 & \textbf{0.603} &
			0.298 \\
			\hline
		\end{tabular}
		
	\end{table}
	\subsubsection{Quantitative  Results:}
	Table~\ref{tab:adaptive_results} summarizes the quantitative comparison between uncertainty-guided adaptive sampling and random sampling. The evaluation includes reconstruction metrics (PSNR, SSIM, and LPIPS) together with uncertainty estimation metrics (Pearson correlation, Spearman correlation, and AUSE) to assess both reconstruction quality and uncertainty reliability.
	\subsubsection{Qualitative Results} 
\begin{figure}[H]
		\centering
		\includegraphics[width=\textwidth]{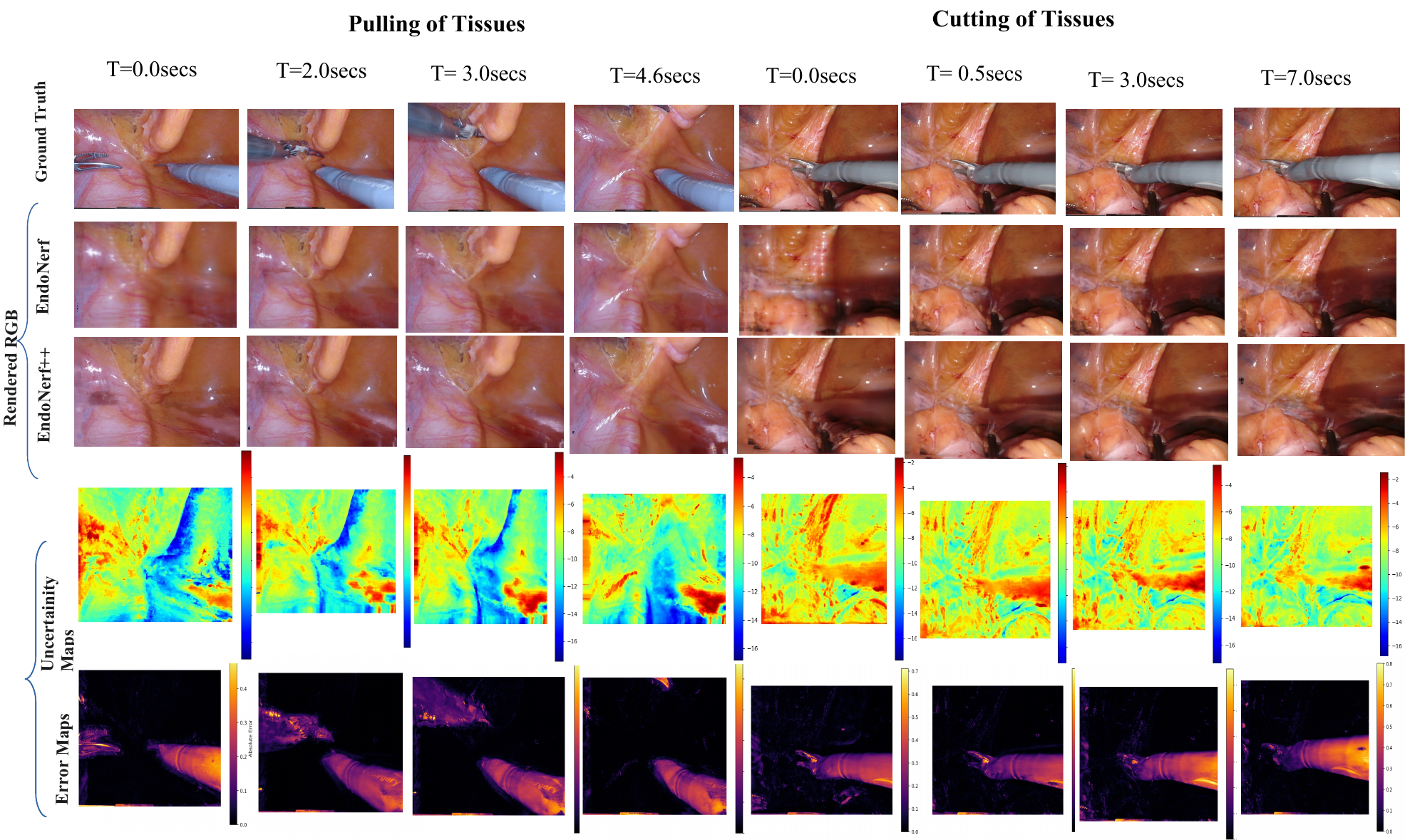}
		\caption{Qualitative comparison of reconstruction results on representative
			frames from the tissue pulling and tissue cutting sequences.}
		\label{fig:adaptive_results}
	\end{figure}

	Figure~\ref{fig:adaptive_results} presents visual comparisons between random sampling and the proposed uncertainty-guided adaptive sampling strategy. By allocating more samples to uncertain regions, the proposed approach produces sharper tissue boundaries, better preserves fine anatomical details, and reduces reconstruction artifacts, demonstrating the practical effectiveness of uncertainty-guided sampling in dynamic surgical scenes.
	Considering both the quantitative and qualitative results, the findings demonstrate that uncertainty-guided adaptive sampling efficiently takes advantage of the anticipated uncertainty to enhance neural rendering. In dynamic surgical settings, the suggested approach provides more accurate uncertainty estimation while achieving competitive reconstruction quality.
	\subsection{Overall Comparison with State-of-the-Art (SOTA)}
	Having evaluated the individual components of the proposed framework, we now assess the overall performance of EndoNeRF++ by comparing it with the state-of-the-art methods EndoNeRF \cite{wang2022endonerf} and EDSSR \cite{long2021edssr}. Quantitative comparisons on the Pulling and Cutting Tissues sequences are summarized in Table~\ref{tab:overall_results}.
	\begin{table}[H]
		\centering
		\caption{Overall Quantitative comparison with SOTA.}
		\label{tab:overall_results}
		
		\footnotesize
		\setlength{\tabcolsep}{3.5pt}
		\renewcommand{\arraystretch}{1.15}
		
		\resizebox{1\textwidth}{!}{
			\begin{tabular}{|l|c|c|c|c|c|c|}
				\toprule
				\multirow{2}{*}{\textbf{Method}}
				& \multicolumn{3}{c|}{\textbf{Pulling Tissues}}
				& \multicolumn{3}{c|}{\textbf{Cutting Tissues}}\\
				
				\cmidrule(lr){2-4}
				\cmidrule(lr){5-7}
				
				& \textbf{PSNR} {\textcolor{ForestGreen}{$\uparrow$}}
				& \textbf{SSIM} {\textcolor{ForestGreen}{$\uparrow$}}
				& \textbf{LPIPS} {\textcolor{BrickRed}{$\downarrow$}}
				& \textbf{PSNR} {\textcolor{ForestGreen}{$\uparrow$}}
				& \textbf{SSIM} {\textcolor{ForestGreen}{$\uparrow$}}
				& \textbf{LPIPS} {\textcolor{BrickRed}{$\downarrow$}}\\
				
				\midrule
				\hline 
				E-DSSR(Baseline)~\cite{long2021edssr}
				& 15.29$\pm$1.287
				& 0.683$\pm$0.042
				& 0.392$\pm$0.038
				& 13.39$\pm$1.387
				& 0.630$\pm$0.057
				& 0.423$\pm$0.047\\
				\hline
				EndoNeRF (Baseline)~\cite{wang2022endonerf}
				& 28.32$\pm$0.503
				& 0.912$\pm$0.015
				& 0.107$\pm$0.019
				& 26.58$\pm$0.520
				& 0.884$\pm$0.012
				& 0.130$\pm$0.020\\
				\hline 
				EndoNeRF++(ours)
				& 29.08$\pm$0.199
				& 0.949$\pm$0.006
				& 0.052$\pm$0.008
				& 27.334$\pm$0.391
				& {0.930}$\pm$0.009
				& {0.076}$\pm$0.008\\
				\hline 
				Unc. EndoNeRF++(ours)
				& 29.256$\pm$0.326
				& 0.943$\pm$0.008
				& 0.055$\pm$0.009
				& 27.356$\pm$0.141
				& 0.922$\pm$0.006
				& 0.087$\pm$0.005\\
				\hline 
				Adap. EndoNeRF++(ours)
				& \textbf{29.537}$\pm$0.551
				& \textbf{0.960}$\pm$0.010
				& \textbf{0.048}$\pm$0.011
				& \textbf{27.377}$\pm$0.646
				& \textbf{0.931}$\pm$0.012
				& \textbf{0.074}$\pm$0.005\\

				\bottomrule
			\end{tabular}
		}
	\end{table}
	
	Table \ref{tab:overall_results} compares the quantitative performance of the proposed methods with EndoNeRF and E-DSSR on the EndoNeRF dataset. On the Pulling Tissues sequence, Adap. EndoNeRF++ achieves the best overall performance, attaining the highest PSNR ($29.537\pm 0.551$) and SSIM ($0.960\pm 0.010$) together with the lowest LPIPS ($0.048\pm 0.011$), demonstrating superior reconstruction fidelity and perceptual quality. On the Cutting Tissues sequence, Adap. EndoNeRF++ achieves the highest PSNR ($27.377\pm 0.646$), whereas EndoNeRF++ provides the best SSIM (0.931±0.012) and LPIPS ($0.074\pm 0.005$), indicating that adaptive uncertainty-guided sampling improves reconstruction accuracy but may not consistently optimize structural similarity and perceptual quality for more challenging tissue deformations.\\ 
	These findings together demonstrate that, the proposed EndoNeRF++ variants consistently outperform the original EndoNeRF and substantially surpass E-DSSR across both sequences, highlighting the effectiveness of the proposed multi-resolution encoding, uncertainty estimation, and adaptive sampling strategies.
	\subsection{Ablation Studies}
	To evaluate the contribution of the suggested components, we conduct ablation investigations. Table \ref{tab:hash_ablation} determines the ideal resolution and scaling factors by evaluating various multi-resolution hash encoding configurations. The progressive performance improvements of EndoNeRF++, Unc. EndoNeRF++, and Adap. EndoNeRF++ shown in Table 3 illustrate the contribution of uncertainty estimation and uncertainty-guided adaptive sampling.
	
	\begin{table}[H]
		\caption{Performance comparison of different multi-resolution hash encoding configurations.}
		\label{tab:hash_ablation}
 	\resizebox{\linewidth}{!}{
			\begin{tabular}{|c|c|l|l|c|l|l|l|}
				\hline
				\textbf{Scenario}                                              & \textbf{Hash Resolution} & \textbf{Scale} & \textbf{Levels} & \textbf{Feature Size} & \textbf{PSNR} {\color{green}$\uparrow$}  & \textbf{SSIM} {\color{green}$\uparrow$} & \textbf{LPIPS} {\color{red}$\downarrow$} \\ \hline
				\multirow{3}{*}{Pulling Tissue}                       & $\{8,16,32\}$                & $\{1.5, 1.8,2.0\}$      & 16       &2              & 28.76     &  0.9362    &  0.0672     \\ \cline{2-8}  & $\{4,16,64\}$ &
				$\{1.5,1.8,2.2\}$ &
				16 & 2 &
				24.4208 &
				0.8703 &
				0.2774     \\ \cline{2-8}  & $\{8,16,32\}$ &
				$\{1.4,1.7,1.9\}$ &
				16 & 8 &
				\textbf{29.0812} &
				\textbf{0.9489} &
				\textbf{0.0524}        \\ \hline
				\multicolumn{1}{|l|}{\multirow{3}{*}{Cutting Tissue}} & $\{8,16,32\}$ &
				$\{1.4,1.7,1.9\}$ &
				16 & 8 &
				\textbf{27.3341} &
				\textbf{0.9302} &
				\textbf{0.0762}       \\ \cline{2-8} 
				\multicolumn{1}{|l|}{} &$\{8,16,32\}$ &
				$\{1.5,1.8,2.0\}$ & 16 & 2 &
				25.6109 & 0.8757 & 0.1479       \\ \cline{2-8} 
				\multicolumn{1}{|l|}{}                                & $\{4,16,64\}$ &
				$\{1.5,1.8,2.2\}$ &
				16 & 2 &
				23.6423 &
				0.8145 &
				0.3085       \\ \hline
		\end{tabular}
}
	\end{table}
	
	The suggested multi-resolution configuration ({8,16,32}, {1.4,1.7,1.9}, 16 levels, feature size 8) presented in Table \ref{tab:hashgrid_results}consistently yields the best reconstruction quality in both tissue deformation situations, according to the results, and is thus used in all subsequent studies. Using this optimal configuration, the quantitative performance of the proposed EndoNeRF++ variants is reported in Table \ref{tab:overall_results}
	\subsection{Cross-Dataset Evaluation}
\begin{figure}[t]
		\centering
		\includegraphics[width=\linewidth]{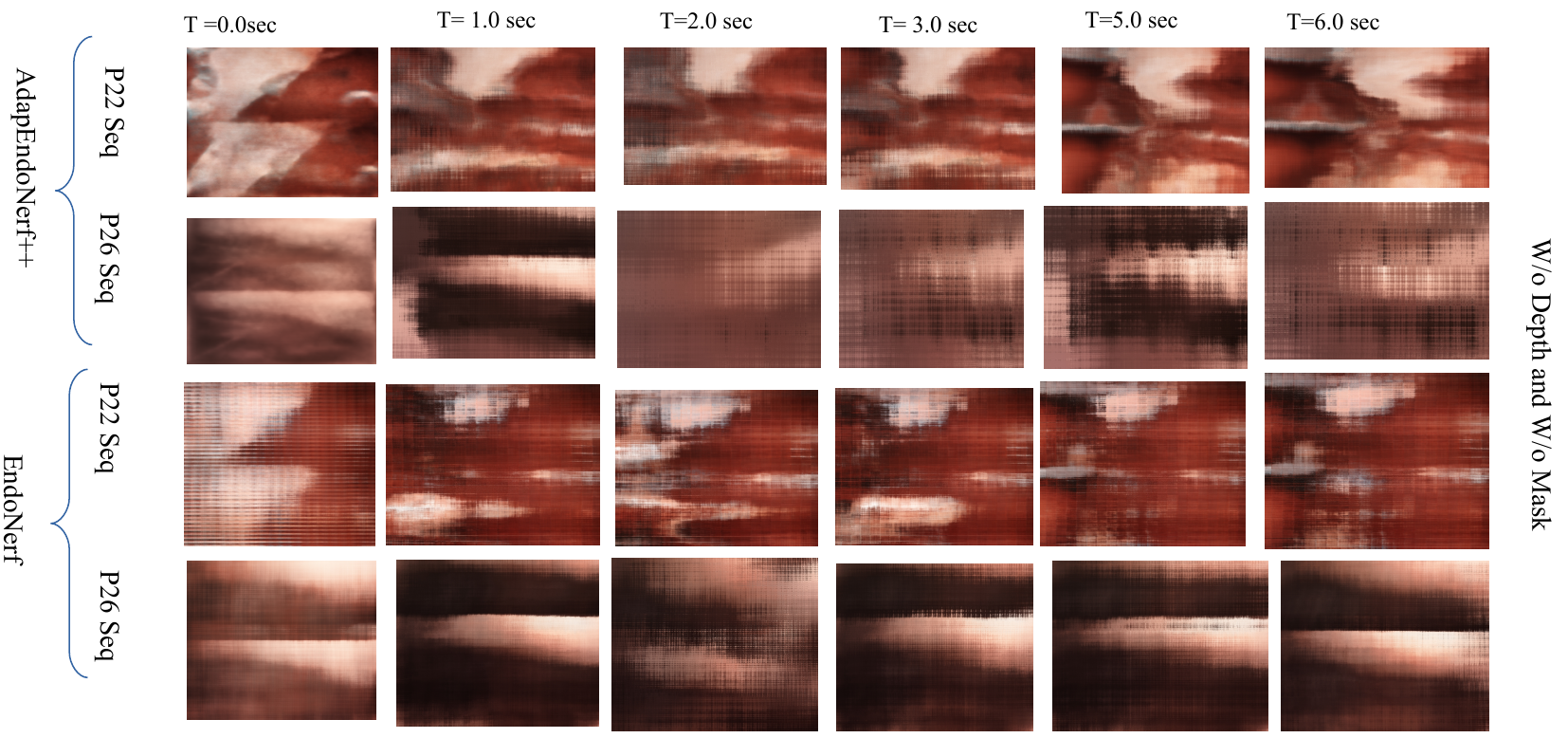}
		\caption{Qualitative comparison on the StereoMIS dataset. Compared with EndoNeRF, the proposed EndoNeRF++ produces sharper tissue boundaries, preserves fine anatomical details, and reduces reconstruction artifacts on both P22 and P26 sequences.}
		\label{fig:stereomis_qualitative}
	\end{figure}

	\textbf{Quantitative  Results:} We validate the suggested framework on the StereoMIS dataset \cite{hayoz2023pose}, which comprises actual robotic surgical scenes obtained with the da Vinci Xi system, in order to assess its resilience and capacity for generalisation. We evaluate the suggested approach using the same evaluation technique and the optimal configuration found from the EndoNeRF trials. Table \ref{tab:stereomis_results} provides a summary of the quantitative findings.
	\begin{table}[t]
		\centering
		\caption{Cross-dataset validation on the StereoMIS dataset. The proposed EndoNeRF++ is compared with the EndoNeRF baseline on the P22 and P26 sequences. Best results are shown in bold.}
		\label{tab:stereomis_results}
		\resizebox{\linewidth}{!}{
			\begin{tabular}{|l|c|c|c|c|c|c|}
				\toprule
				\multirow{2}{*}{\textbf{Model}} &
				\multicolumn{3}{c|}{\textbf{P22 Seq}} &
				\multicolumn{3}{c}{\textbf{P26 Seq}} \\
				\cmidrule(lr){2-4}\cmidrule(l){5-7}
				& \textbf{PSNR} {\color{green}$\uparrow$}
				& \textbf{SSIM} {\color{green}$\uparrow$}
				& \textbf{LPIPS} {\color{red}$\downarrow$}
				& \textbf{PSNR} {\color{green}$\uparrow$}
				& \textbf{SSIM} {\color{green}$\uparrow$}
				& \textbf{LPIPS} {\color{red}$\downarrow$} \\
				\midrule
				AdapEndoNeRF++ &
				\textbf{11.021 $\pm$ 1.222} &
				\textbf{0.319 $\pm$ 0.036} &
				\textbf{0.719 $\pm$ 0.040} &
				\textbf{11.195 $\pm$ 1.905} &
				\textbf{0.294 $\pm$ 0.083} &
				\textbf{0.612 $\pm$ 0.047} \\
				\midrule
				EndoNeRF &
				10.294 $\pm$ 1.561 &
				0.257 $\pm$ 0.041 &
				0.741 $\pm$ 0.057 &
				9.930 $\pm$ 1.200 &
				0.283 $\pm$ 0.076 &
				0.695 $\pm$ 0.050 \\
				\bottomrule
		\end{tabular}}
	\end{table}
	The proposed EndoNeRF++ consistently performs better than the EndoNeRF baseline on both P22 and P26 sequences, as Table \ref{tab:stereomis_results} illustrates. These findings show that the suggested approach improves reconstruction fidelity while successfully generalising to previously encountered robotic surgical situations.\\
	\textbf{Qualtative Results:} A qualitative comparison of EndoNeRF and the suggested EndoNeRF++ on the StereoMIS dataset is shown in Figure \ref{fig:stereomis_qualitative}. The quantitative results shown in Table \ref{tab:stereomis_results} are supported by the suggested approach, which produces sharper reconstructions with better structural consistency and less visual artefacts.
	The qualitative comparisons, which produce visually sharper tissue reconstruction and improved visual clarity on both StereoMIS sequences, further validate the superiority of the suggested EndoNeRF++.
	\section{Conclusion, Limitations and Future Work}
	This paper introduced Endo-NeRF++, an uncertainty-aware neural rendering framework designed for the reconstruction of dynamic surgical scenes. The proposed method enhances the quality of reconstruction and the consistency of temporal information in deformable surgical scenes by incorporating adaptive uncertainty-aware sampling, temporal feature merging, and multi-resolution hash-grid encoding.\\
	The experimental results of robotic surgical sequences showed that the reconstruction fidelity and rendering consistency were enhanced for dynamic tissue interactions, such as pulling and cutting of tissues.\\
	The study is constrained by the scarcity of surgical datasets, restricted perspectives, and distorted endoscopic observations resulting from occlusions, reflections, and tissue deformation. Moreover, neural rendering techniques continue to demand high computational power and memory, which poses significant challenges for their use in real-time systems with constrained resources.\\
	Future research will aim to minimize the computational and memory demands of NeRF-based approaches, enhance scalability and real-time capability, and broaden the framework to accommodate larger multi-view surgical datasets and more reliable uncertainty estimation methods.\\
	\textbf{Data Availability:} The data will be made available upon reasonable request through the
	\href{https://docs.google.com/forms/d/e/1FAIpQLSfM0ukpixJkZzlK1G3QSA7CMCoOJMFFdHm5ltCV1K6GNVb3nQ/viewform}{data-request form}.
	\bibliographystyle{splncs04}
	\bibliography{ref}
\end{document}